\documentclass[aps,prb,twocolumn,showpacs,floatfix]{revtex4}
\usepackage{amssymb}
\usepackage{graphicx}

\begin{document}

\title{Spontaneous symmetry breaking of magnetostriction
in metals with multivalley band structure}

\author{G.~P.~Mikitik}
\affiliation{B.~Verkin Institute for Low Temperature Physics \&
Engineering, Ukrainian Academy of Sciences, Kharkov 61103,
Ukraine}

\author{Yu.~V.~Sharlai}
\affiliation{B.~Verkin Institute for Low Temperature Physics \&
Engineering, Ukrainian Academy of Sciences, Kharkov 61103,
Ukraine}

\begin{abstract} We show that a first-order
phase transition can take place in a metal in a strong magnetic
field if an electron Landau level approaches the Fermi energy of
the metal. This transition is due to the electron-phonon
interaction and is characterized by a jump in magnetostriction of
the metal. If there are several equivalent groups of charge
carriers in the metal, a spontaneous symmetry breaking of the
magnetostriction can occur when the Landau level crosses the Fermi
energy, and this breaking manifests itself as a series of the
structural phase transitions that change a crystal symmetry of the
metal. With these results, we discuss unusual findings recently
discovered in bismuth.
\end{abstract}


\maketitle

\section{Introduction}

Recently, \cite{B1} oscillations of the Nernst coefficient in
bismuth were observed for the magnetic fields directed along the
trigonal axis of the crystal. These oscillations have the shape of
peaks that originate from the crossing of the Landau levels of
the electrons and holes in bismuth with the Fermi level $\mu$ of
this semimetal. \cite{SM1,AB,SM2} However, several unusual peaks
of this coefficient were also discovered for very high magnetic
fields $H$ ($14\lesssim H\lesssim 33$ T). \cite{B2,KBstrangePeaks}
At such magnetic fields almost all the Landau levels are empty,
and the unusual peaks cannot result from the above-mentioned
crossing. In this context Behnia {\it et al}. \cite{B2} suggested
that the unusual peaks are caused by some collective effects in
the electron system of bismuth. Interestingly, in the same
interval of the magnetic fields directed almost along the trigonal
axis, jumps of magnetization were observed which were ascribed to
field-induced instabilities of the ground state of interacting
electrons in bismuth. \cite{Ong} Various explanations of the
unusual peaks were put forward. \cite{Ser,B3,Ser1,Zhu} In
particular, the recent study of their angular variation with a
rotating magnetic field led to the conclusion that they are
produced by the presence of a secondary domain in twinned
crystals. \cite{Zhu} However, this scenario leaves a number of
questions unanswered, \cite{ZhuQuestions} and it does not explain
the observation of hysteretic jumps in magnetization. \cite{Ong}

It is well known that crystals are deformed in a magnetic field,
i.e., they exhibit magnetostriction. \cite{Sh} Similarly to the de
Haas - van Alphen effect, the magnetostriction oscillates with
changing magnetic field. In this paper we show that apart from the
oscillations, jumps in the magnetostriction can occur when the
Landau levels approach the Fermi level $\mu$ of a metal. These
first-order phase transitions can take place if there are, at
least, two different groups of charge carriers in the metal. For
example, this situation occurs in bismuth in which the Fermi
surface consists of the electron and hole parts. Moreover, in
bismuth the electron part is composed of three equivalent
ellipsoids. When the magnetic field is along the trigonal axis of
bismuth, one may expect that the deformation of the crystal does
not destroy its symmetry. However, we show in this paper that if a
Landau level of equivalent electron pockets in a metal with a
multivalley band structure is close to the Fermi energy $\mu$, a
spontaneous symmetry breaking of the crystal deformation can occur
so that the electron pockets become nonequivalent. In other words,
with increasing $H$, the Landau level of the different pockets crosses the Fermi energy at different $H$, and we finds several phase transitions instead of single one. These first-order phase
transitions change the crystal symmetry of the metal in a certain
interval of magnetic fields. Thus, in fact, we find that there is
a possibility of governing the crystal symmetry in multivalley
metals with the magnetic field. Interestingly, in the recent
experimental investigations \cite{behnia11} of the magnetoresistivity of bismuth and of its magnetostriction measured
\cite{behnia} for magnetic fields near the trigonal axis, unusual
angular asymmetries of these quantities were observed when Landau
levels of appropriate electron ellipsoids were close to the Fermi
energy. We discuss how the predicted phase transitions can explain the effects of
Refs.~\onlinecite{B2}-\onlinecite{Ong,behnia11,behnia}.

In principle, spontaneous symmetry breaking of equivalent electron
groups in a multivalley metal can be due to the electron-electron
interaction. \cite{abanin} In this paper we show that  the effect
of the spontaneous symmetry breaking can also result from the
electron-phonon interaction inducing the magnetostriction, and
this effect can occur even without the interaction between the
electrons. Because of this, to simplify our analysis, we neglect
the electron-electron interaction here. The relative role of the
electron-phonon and electron-electron interactions in the symmetry
breaking should be analyzed for every specific metal
separately. We also note that the spontaneous symmetry breaking of
the magnetostriction is reminiscent the Jahn-Teller effect.
\cite{JT} However, as we shall see, there is a difference between  these effects.

The paper is structured as follows: In Sec.~\ref{I} we outline simple considerations that qualitatively explain the origin of the transitions and the spontaneous symmetry breaking. These considerations precede the extended quantitative analysis given in the subsequent sections. In Sec.~\ref{II} general formulas for the magnetostriction of a multivalley metal are presented. To clarify the essence of the matter, in Sec.~\ref{III} we calculate the magnetostriction using the simplest model for the band structure of the metal. In Sec.~\ref{IV} the effect of the spontaneous symmetry breaking is discussed. We first consider this effect for a model imitating the band structure of bismuth, and then discuss a generalization of the obtained results to an arbitrary metal with the equivalent groups of charge carriers. In Sec.~\ref{V} the conclusions are presented, and the Appendix contains some details of the calculations.

\section{Qualitative considerations}\label{I}

On switching a magnetic field on, the conduction electrons of metals fill the Landau levels (the Landau subbands), and their energy changes as compared to the energy without the field. Due to the electron-phonon interaction, this change in the energy produces an elastic deformation of a metal, i.e, its magnetostriction. This  deformation shifts the electron energy bands and thereby the edges of the Landau subbands in proportion to the magnitude of the deformation. As a result, in the deformed crystal the total energy consisting of the electron and elastic parts reaches its minimum.  This principle of the energy minimization just determines the magnitude and the type of the deformation. Evidently, under this deformation a gain in the electron energy exceeds the loss in the elastic energy.

For simplicity, let us assume that the Fermi energy of the electrons in the metal is independent of the deformations and the magnetic field. When the magnetic field increases (and we neglect the magnetostriction), the Landau levels gradually cross this Fermi energy. However, if the magnetostriction is taken into account, the crossing occurs in the form of small jumps of the Landau levels, and at such jumps the system consisting of the electrons and of the crystal deformations goes from one minimum of its total energy to another minimum. To explain this statement, consider the situation when the edge of the $l$-th Landau subband in the deformed crystal lies slightly above the Fermi energy. In this state of the crystal the total energy is at its minimum. If an additional deformation appears, and the  magnitude $u$ of this deformation is such that the $l$-th subband begins to fill with the electrons, the energy of the electrons occupying this Landau level can be estimated as  $\Delta_l\cdot N^{(l)}(H)$ where we measure the electron energies from the Fermi level, $\Delta_l$ defines the edge  of the $l$-th subband ($\Delta_l<0$), and  $N^{(l)}(H)$ is the number of the electrons in this subband, $N^{(l)}(H)\propto H\sqrt{|\Delta_l|}$. \cite{Sh} On the other hand, the loss in the elastic energy at this deformation  will be  proportional to $u^2\propto [\Delta_l -\Delta_l(0)]^2$ where $\Delta_l(0)>0$ is the initial value of $\Delta_l$ at $u=0$. Since at sufficiently small $\Delta_l$ and $\Delta_l(0)$, the gain in the electron energy  $|\Delta_l|\cdot N^{(l)}(H)\propto |\Delta_l|^{3/2}$ always exceeds the elastic term quadratic in $\Delta_l$, a deeper minimum of the total energy than the initial one exists, and the initial state of the crystal is not optimal at such small values of $\Delta_l(0)$. It is favorable for the deformation to make a jump and for the electrons to occupy partly the $l$-th Landau level at a certain small value of $\Delta_l(0)$. This jump means that a first-order phase transition occurs in the metal.

The above  considerations are valid for any type of the deformation. In reality, that type of the deformations occurs which provides the deepest minimum of the total energy. In metals with several equivalent groups of charge carriers, only a  part of elastic deformations leaves the symmetry of the crystal unchanged, whereas the other part breaks it. For the symmetry breaking to occur, an asymmetric deformation should provide a deeper minimum of the total energy than any symmetric deformation. This condition imposes some restriction on the elastic moduli of the metal and the components of the deformation potential (i.e. on the constants defining the ratios $[\Delta_l-\Delta_l(0)]/u$ for different types of the deformations). Although the explicit form of this restriction depends on the crystal symmetry, our analysis shows that the restriction is not rigid and may be fulfilled for many metals including bismuth, see Sec.~\ref{IV} and the Appendix.

\section{Formulas for magnetostriction}\label{II}

At low temperatures the magnetostriction can be found from the
minimization of the following energy $E$ with respect to the
deformation $u$:
 \begin{eqnarray}\label{1a}
E(u,H)=C\frac{u^2}{2} +\Delta
E_{e}(u,H)-\Delta E_{e}(u,0),
 \end{eqnarray}
where
\[
 \Delta E_{e}(u,H)\equiv E_e(u,H)- E_e(0,H),
 \]
$E_e=\sum_{i=1}^{n}E_i$ is the electron energy in a multivalley
metal with $n$ groups of charge carriers, $E_i$ is the energy of
the $i$-th group, $u$ is the magnitude of the deformation tensor,
and $C$ is the appropriate elastic modulus of the crystal. The
first term in Eq.~(\ref{1a}) gives the total elastic energy
$Cu^2/2$ of the deformation. This energy is partly produced by  $\Delta E_{e}(u,0)$, and hence the difference of the first and third
terms is the elastic energy that is not associated with the
electron groups under study. The second term describes the
change in the energy of these groups in the magnetic
field under the deformation. Here we use the simplified form
$Cu^2/2$ of the elastic energy. Its accurate form should take into
account a symmetry of the metal; see Sec.~\ref{IV}.

The calculation of the magnetostriction $u$ with formula
(\ref{1a}) is in agreement with the traditional approach presented
in Ref.~\onlinecite{Sh}. Indeed, the difference $\Delta
E_{e}(u,H)-\Delta E_{e}(u,0)$ can be rewritten in the form:
\begin{eqnarray*}
\Delta E_{e}(u,H)-\Delta
E_{e}(u,0)\!\!\!&=&\!\!\![E_e(u,H)-E_e(u,0)] \nonumber
\\-[E_{e}(0,H)-E_{e}(0,0)]\!\!\!&=&\!\!\!\!-
 \!\!\int_0^H\!\!\!\! [M_e(u,H')\!-\!\!M_e(0,H')]dH'\!,
\end{eqnarray*}
where $M_e(u,H)$ is the magnetization of the $n$ electron groups
in the crystal under the deformation $u$ and at the magnetic field
$H$. Assuming that $M_e(u,H)$ is a smooth function of
$u$, i.e., $M_e(u,H')-M_e(0,H')\approx u\cdot [\partial
M_e(u,H')/\partial u]_{u=0}$, the minimization of expression
(\ref{1a}) in $u$ gives
\begin{eqnarray}\label{1b}
u=\frac{1}{C}\int_0^H\! \left[\frac{\partial M_e(u,H')}
{\partial u}\right]_{u=0}dH'.
\end{eqnarray}
At weak magnetic fields  when $M_e=\chi_e H$, we arrive at the
quadratic dependence of $u$ on $H$:\cite{Kap}
\begin{eqnarray}\label{1c}
u=\frac{H^2}{2C}\frac{\partial \chi_e}{\partial u}|_{u=0},
\end{eqnarray}
where $\chi_e$ is the magnetic susceptibility of the electron
groups involved. With increasing magnetic field, the
oscillating part of the magnetization, $M_e^{\rm os}$, appears, and
if only the main harmonics of this oscillating part are taken into
account,
\[
M_e^{\rm os}\approx \sum_{i=1}^{n}M_i^{\rm os}= \sum_{i=1}^{n}
 M_i\sin(\frac{2\pi S_i}{H}+\phi_i),
\]
one obtains the following formula that agrees with expression (4.20) from Ref.~\onlinecite{Sh}:
\begin{eqnarray*}
u\approx -\frac{H}{C}\sum_{i=1}^{n}\left[\frac{\partial \ln
S_i}{\partial u}M_i^{\rm os}\right]_{u=0},
\end{eqnarray*}
where $S_i$ are extremal cross-section areas of the Fermi surfaces
of the electron groups, and $M_i$, $\phi_i$ are the magnitudes and
phases of the harmonics. Below we shall deal with the expression
(\ref{1a}) rather than with formula (\ref{1b}) since we shall take
into account a {\it nonanalytic} contribution (in relation to the
variable $u$) to the electron energy. It is this contribution that
determine the effects discussed in this paper.

The differences $\Delta E_{e}$ caused by the deformation originate
from the changes in the energy spectra $\epsilon_{i}({\bf k})$ for
the electron groups. These changes $\Delta\epsilon_{i}({\bf k})$
can be described with the deformation potential $D({\bf k})$,
$\Delta\epsilon_{i}({\bf k}) =D_{i}({\bf k})u$. For simplicity, we
assume below that $D_{i}({\bf k})$ are constants which are
independent of ${\bf k}$, i.e., the deformation shifts the
electron bands as a whole by the values $\Delta\varepsilon_{i}=
D_{i}u$. An analysis of $D_i$ for bismuth \cite{Han} shows that
this ``solid-band'' approximation is really  good.

Strictly speaking, formula (\ref{1a}) has to include also a
deformation change in the energy of the bands completely
filled with electrons, and accordingly expression (\ref{1b}) contains the total magnetization $M$ rather than $M_e$. However, the filled bands give an analytic and nonoscillating contribution to $M$ in the
variables $H$ and $u$, and the deformation change of the magnetic
energy for these bands is $-uH^2(\partial \chi_{\rm
filled}/\partial u)_{u=0}/2$. Since the magnetic susceptibility
$\chi_{\rm filled}$ of the filled bands lying well below the Fermi
level $\mu$ weakly depends on the deformation $u$ (in the
solid-band approximation the magnetic susceptibility of the filled
bands is completely independent of $u$), these bands can lead only
to a small renormalization of the coefficient $\beta$ in
Eq.~(\ref{7a}), and we  completely neglect these bands here.

\section{Simple model}\label{III}

For a metal with a single electron group we obtain $\Delta E_{e}(u,H)=\Delta E_{e}(u,0)$ within the solid-band approximation, and the minimization of $E(u,H)$ in $u$ gives $u=0$. In other words, within this approximation the magnetostriction is absent for
such a metal. To clarify the main ideas, consider the simplest
model which leads to a nonzero value of the magnetostriction. In
this model the Fermi surface of a metal consists of two spheres
corresponding to small and large electron groups with the spectra
\[
\epsilon_{s,l}({\bf k})=\varepsilon_{s,l}+\frac{\hbar^2{\bf k}^2}
{2m_{s,l}}.
\]
Here $\varepsilon_{s,l}$ are the energy minima of the electron
groups with $\mu-\varepsilon_{s}\ll \mu-\varepsilon_{l}$, ${\bf
k}$ is the wave vector in the Brillouin zone of the crystal (for
each group  ${\bf k}$ is measured from the appropriate minimum),
$m_{s,l}$ are the effective masses, and $\mu$ is the Fermi level
of the metal. The large difference in the sizes of the groups,
$\mu- \varepsilon_{s} \ll \mu-\varepsilon_{l}$, is exclusively
assumed to simplify the calculation of the Fermi energy $\mu$
which generally depends on $H$ and $u$. These dependences of $\mu$ can be always represented in the form: $\mu=\mu(H)+\Delta
\mu(u,H)$ where $\mu(H)$ is the Fermi energy at $u=0$, and
$\Delta \mu(u,H)$ describes the  dependence of $\mu$ on $u$. These
$\mu(H)$ and $\Delta \mu(u,H)$ are found from the conservation of the electrons, $N_{s}+ N_{l}=N_{tot}=$ const., where $N_{s,l}$ are the
numbers of the particles in the small and large groups, and
$N_{tot}$ is the total number of the electrons. When $N_l\gg N_s$,
this conservation law leads to simple formulas: $\mu(H)\approx \mu(0)$ and $\Delta
\mu(u,H)\approx \Delta \mu(u)\approx \Delta\varepsilon_{l}=D_lu$.

Within the solid-band approximation, one may write at $H=0$,
\begin{eqnarray}\label{2a}
\Delta E_e(u,0)=\Delta\varepsilon_{s}N_{s}(0) +
\Delta\varepsilon_{l}N_{l}(0),
\end{eqnarray}
where $N_{s,l}(0)\equiv N_{s,l}(\mu-\varepsilon_{s,l},H=0)$ are
the numbers of the particles in the groups at $H=0$ and $u=0$,
 \[
 N_{s,l}(0)=\frac{(2m_{s,l})^{3/2}(\mu-\varepsilon_{s,l})^{3/2}}{
3\pi^2\hbar^3}.
 \]
On closer examination of $\Delta E_e(u,0)$ one should take into account a change of $N_{s,l}$ in the process of a variation of $u$ (i.e., of $\Delta\varepsilon_{s,l}=D_{s,l}u$). Then the right hand side of Eq.~(\ref{2a}) is rewritten as follows:
 \begin{eqnarray}\label{3a}
&\Delta&\!\!\! E_{e}=\int_{\varepsilon_{s}}^{\varepsilon_{s} +
\Delta\varepsilon_{s}}\!\!\!\!\!\!\!\!N_{s}(\mu'-\varepsilon_{s}',0)
d\varepsilon_{s}'+\!\!\int_{\varepsilon_{l}}^{\varepsilon_{l} +
\Delta\varepsilon_{l}}\!\!\!\!\!\!\!\!N_{l}(\mu'-\varepsilon_{l}',0)
d\varepsilon_{l}'\nonumber \\
&=&\!\!\!N_{tot}\Delta\varepsilon_{l}+(1-\frac{D_l}{D_s})
\int_{\varepsilon_{s}}^{\varepsilon_{s}+
\Delta\varepsilon_{s}}\!\!\!\!\!\!\!\!N_{s}(\mu'-\varepsilon_{s}',0)
d\varepsilon_{s}' \nonumber \\
&=&\!\!\!N_{tot}\Delta\varepsilon_{l}+
\Omega_{s}(\mu\!\!+\!\!\Delta \mu\!\!-
\!\!\Delta\varepsilon_{s}\!\!-\!\! \varepsilon_{s},0)\!\!
-\!\!\Omega_{s}(\mu\!\!-\!\!\varepsilon_{s},0),\ \ \ \
 \end{eqnarray}
where $\Omega_{s}(\mu'-\varepsilon_{s}',0)$ is the $\Omega$
potential of the small group, and $N_{s}(\mu'-
\varepsilon_{s}',0)$, $N_{l}(\mu'-\varepsilon_{l}',0)$ denote the
numbers of the particles in the electron groups at $H=0$ provided  that the energy minima of these groups are equal to
$\varepsilon_{s}'$ and to $\varepsilon_{l}'$, respectively, and
$\mu'=\mu'(\varepsilon_{s}')$. In obtaining Eq.~(\ref{3a}), we
have used the conservation law $N_l=N_{tot}-N_s$ and the
equalities
\[
N_{s}(\mu-\varepsilon_{s},0)=-\frac{\partial \Omega_{s}}{\partial
\mu}=\frac{d \Omega_{s}}{d
\varepsilon_{s}}(1-\frac{D_l}{D_s})^{-1}
 \]
where
\[
\Omega_{s}(\mu-\varepsilon_{s},0)=- \frac{2(2m_{s})^{3/2}
(\mu-\varepsilon_{s})^{5/2}}{15\pi^2\hbar^3}
 \]
for  $\mu-\varepsilon_{s}>0$, and $\Omega_{s}(\mu-
\varepsilon_{s},0)=0$  for $\mu-\varepsilon_{s}<0$.  Taking into
account the relationship $\Delta\mu- \Delta\varepsilon_{s}
=(D_l-D_s)u$, we see that $\Omega_{s}(\mu+\Delta \mu-
\Delta\varepsilon_{s} -\varepsilon_{s},0)$ in Eq.~(\ref{3a}) is a
{\it nonlinear and nonanalytic} function of $u$ when in the deformed crystal the band bottom approaches the Fermi level, i.e., when  \[\Delta \mu- \Delta\varepsilon_{s} +\mu-\varepsilon_s \to 0.
 \]
However, in the opposite limit, at relatively small shifts of $\varepsilon_s$ and $\mu$
 \begin{equation}\label{ineq}
 |\Delta\mu- \Delta\varepsilon_{s}|\ll \mu(0)- \varepsilon_{s},
 \end{equation}
this $\Omega_{s}(\mu+\Delta \mu- \Delta\varepsilon_{s}
-\varepsilon_{s},0)$  tends to a linear function of $u$,
 \[
 \Omega_{s}(\mu+\Delta \mu-
\Delta\varepsilon_{s}-\varepsilon_{s},0)\approx \Omega_{s}(\mu-
\varepsilon_{s},0)-N_s(0)(\Delta \mu- \Delta\varepsilon_{s}),
 \]
and Eq.~(\ref{3a}) reduces to Eq.~(\ref{2a}). Our estimates show that under the condition $N_l\gg N_s$ the inequality (\ref{ineq}) is
always fulfilled {\it at zero magnetic field}, and hence
Eq.~(\ref{2a}) is valid in this case. However, another situation
occurs at $H\neq 0$.

In the magnetic field $H$, the electrons fill the Landau levels
(the Landau subbands) of both the groups,
 \[
 \epsilon_{s,l}^n(k_z)=\varepsilon_{s,l}+\frac{\hbar
eH}{m_{s,l}c}(n+\frac{1}{2})+ \frac{\hbar^2k_z^2}{2m_{s,l}},
 \]
where $e$ is the absolute value of the electron charge, $n=0,1,
\dots$, and $k_z$ is directed along the magnetic field. For
simplicity, we neglect the intrinsic magnetic moment of an
electron  here. Let the edge $\epsilon_s^1(0)$ of the first
Landau subband of the small group be in the vicinity of $\mu$.
This occurs at $H\approx H_1\equiv (2/3)(m_sc/\hbar
e)(\mu-\varepsilon_s)$. The calculation of $\Delta E_e(u,H)$ is
similar to the calculation of $\Delta E_e(u,0)$, and we obtain
\begin{eqnarray}\label{5a}
\Delta E_e(u,H)=\Delta\varepsilon_{l}N_{tot}+
\Delta\varepsilon_{s}N_{s}^0(H)\left(\!\!1\!\!-\!\!
\frac{D_l}{D_s}\right) \nonumber \\ +
\Omega_{s}^1(\mu\!\!+\!\!\Delta
\mu\!\!-\!\!\Delta\varepsilon_{s}\!\!-\!\!
\epsilon_{s}^1(0),H\!)\!\!-\!\!\Omega_{s}^1(\mu\!\!-
\!\!\epsilon_{s}^1(0),H) ,
\end{eqnarray}
where $N_{tot}$ is the total number of the electrons in the two
groups; $N_s^0(H)$ is the number of the particles in the zeroth
Landau level of the small group at $u=0$,
 \[
 N_s^0(H)=\frac{eH\sqrt{2m_s}[\mu-\epsilon_s^0(0)]^{1/2}}{\pi^2
\hbar^2 c},
 \]
and $\Omega_{s}^1(\mu
-\epsilon_{s}^1(0),H)$ is the $\Omega$ potential of the electrons
in the first Landau level of this group,
 \begin{eqnarray}\label{6a}
\Omega_{s}^1(\mu\!\!-\!\!\epsilon_{s}^1(0),H)\!=
-\frac{2eH\sqrt{2m_s}
[\mu\!-\!\epsilon_s^1(0)]^{3/2}}{3\pi^2\hbar^2 c}.
 \end{eqnarray}
Now $\Delta\mu- \Delta\varepsilon_{s}$ may be comparable
with $\mu-\epsilon_s^1(0)$, and so we do not replace the last
terms in Eq.~(\ref{5a}) by $(\Delta\varepsilon_{s}-\Delta
\mu)N_s^1(H)$ where $N_s^1(H)$ is the number of the electrons in
the first Landau level of the small group at $u=0$. It is also
necessary to emphasize that formula for $\Delta E_e(u,H)$ does not
reveal a nonlinear and nonanalytic term at the magnetic fields
when a Landau-subband edge of the large group is close to $\mu$.
This is due to our approximation $\Delta\mu= \Delta\varepsilon_l$
based on a small ratio of the densities of states for the small
and large groups. If $N_l$ and $N_s$ were comparable, the
nonanalytic term would also appear when Landau-subband edges of
the large group cross $\mu$.

Combining formulas (\ref{1a}), (\ref{2a}), (\ref{5a}), (\ref{6a}),
we arrive at
\begin{eqnarray}\label{7a}
E(u,H)=C\frac{u^2}{2}+\beta u +\alpha [\Delta_1 -\Delta
D\cdot u]^{3/2}+E_1,
\end{eqnarray}
where the constant $E_1=-\Omega_{s}^1(\mu-\epsilon_{s}^1(0),H)$ is
independent of $u$; $\Delta D\equiv D_s-D_l$;
 \begin{eqnarray}
\Delta_1=\mu-\epsilon_s^1(0)= (\mu-\varepsilon_{s})\frac{H_1
-H}{H_1};\label{9} \\ \alpha(H)\approx \alpha(H_1)=
-\frac{2eH_1\sqrt{2m_s}}{3\pi^2\hbar^2c}; \label{10} \\
\beta(H)\approx \beta(H_1)= [N_s^0(H_1)-N_s(0)]\Delta D; \label{11}
 \end{eqnarray}
the singular (nonanalytic) term $\alpha [\Delta_1 -\Delta
D\cdot u]^{3/2}$ exists only at $\Delta_1 -\Delta D\cdot u>0$, otherwise it is zero. Formula (\ref{7a}) without the singular term is analogous to expressions used in Ref.~\onlinecite{Mi}. The singular term was taken into account in Ref.~\onlinecite{Littlewood}. However, the sign of the parameter $\alpha$ was positive in that paper, whereas we obtain the negative $\alpha$. \cite{com}

 \begin{figure}[t] 
\includegraphics{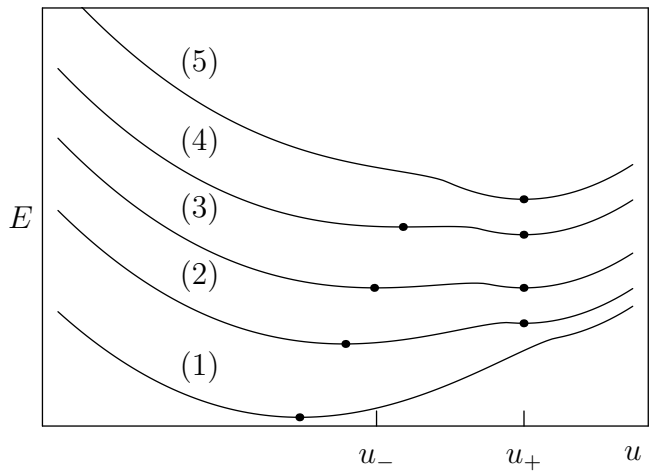}
\caption{\label{fig1} The $u$-dependence of $E$, Eq.~(\ref{7a}),  shown schematically for different $H$: (1)
$H<H_1^-$; (2) $H_1^-<H<H_t$; (3) $H=H_t$; (4) $H_t<H<H_1^+$; (5)
$H_1^+<H$. The points mark the appropriate minima. For clarity,
the curves are shifted along the $E$ axis.
 } \end{figure}   

Our analysis of Eq.~(\ref{7a}) shows that in contrast to the case
$\alpha>0$, at negative $\alpha$ the function $E(u,H)$  has two
minima with respect to $u$ in the interval of the magnetic
fields, $H_1^-<H<H_1^+$, near $H_1$, see Fig.~\ref{fig1}. Here the
boundaries of the interval $H_1^-$ and $H_1^+$ are determined by
the formulas:
\begin{eqnarray}
\frac{H_1\!-\!H_1^-}{H_1}\!&=&-\frac{\beta(H_1)\Delta
D}{C(\mu-\varepsilon_s)};\label{12} \\
\frac{H_1^+\!-\!H_1^-}{H_1}\!&=&\frac{9\alpha^2\Delta D^4}{16
C^2(\mu-\varepsilon_s)}. \label{13}
\end{eqnarray}
In this interval one of the minima occurs at $u=u^+=-\beta/C$,
whereas the second minimum is at the point $u=u^-$ which depends
on $H$. The appearance of this additional minimum is due to
existence of the singular term which is proportional to
$-[(\Delta_1/\Delta D)-u]^{3/2}$ for positive $[(\Delta_1/\Delta
D)-u]$. This term overcomes the quadratic dependence $C(u-u^+)^2/2$
of $E$ that would occur in the vicinity of the point $u=u^+$ if
the singular term were absent. At small positive $H-H_1^-$ the
absolute minimum of $E$ is at $u=u^-$. With increasing $H$, the second minimum at $u=u^+$ decreases, and at the
magnetic field $H_t=(H_1^-+3H_1^+)/4$ one finds $E(u^+)=E(u_0^-)$
where
 \[
 u_0^-\equiv u^-(H_t)=u^+-\frac{27\alpha^2\Delta D^3}{16C^2}.
 \]
At this $H=H_t$ the deformation $u$ jumps from $u_0^-$ to $u^+$ since the absolute minimum of $E$ is at $u=u^+$ for $H>H_t$. Thus, we find a first-order phase transition at the magnetic field $H_t$.

At the first-order phase transition described here, the first
Landau-level edge of the small group sharply crosses the Fermi
energy, and the magnetic moment $M$ associated with this group
experiences the jump,
\begin{equation}\label{14}
 \Delta M \approx \frac{9\sqrt{2m_s}\alpha e(\Delta
D)^2(\mu-\varepsilon_s)}{8\pi^2c\hbar^2 C}.
\end{equation}
Note that under a cycling of $H$, a hysteresis of the magnetic moment and the magnetostriction can occur, and the width of the hysteresis loop may reach $H_1^+-H_1^-$.

The derived jump and hysteresis of $M$ can qualitatively explain
the results of Ref.~\onlinecite{Ong}, in which the magnetization of bismuth was measured at the magnetic fields $H$ tilted away from the
trigonal axis of the crystal by angles $\theta$. For such tilted $H$ the electron ellipsoids in bismuth are not equivalent, and the jump in $M$ just occurs when the Landau-level edge of one of the ellipsoids is close to the Fermi energy (cf. Figs.~3 and 3a in
Refs.~\onlinecite{Ong} and \onlinecite{SM1}, respectively). This
is in accordance with our results if we consider the other
electrons and the holes in bismuth as the large group of charge
carries.

Simple generalization of the above results shows that the
described phase transition can occur in the vicinities of the
magnetic fields
 \[
 H_l\equiv \frac{2(\mu-\varepsilon_s)m_sc}{(2l+1)\hbar e}=
 \frac{3H_1}{2l+1}
 \]
when the $l$-th Landau-level edge of the
small group is close to the Fermi energy (here $l=0,1,2,\dots$).
In this case formula (\ref{7a}) is still valid, the parameter
$\alpha$ is proportional to $H_l$, and the jump $(27\alpha^2\Delta
D^3/16C^2)$ in $u$ at the transition decreases like $(2l+1)^{-2}$
with increasing $l$. The characteristic region of the transition,
$H_l^+\!-\!H_l^-$, also decreases with $l$: $H_l^+\!-\!H_l^-
\propto (2l+1)^{-3}$. Therefore, one may expect that at high $l$ the
transition is smeared by temperature, and so it can be probably
observed only for not-too-large $l$, i.e., at sufficiently strong
magnetic fields, and at low temperatures.

\section{Spontaneous symmetry breaking}\label{IV}

In  a metal each component of the deformation tensor $u_{ij}$, in general, has an effect on its electron spectrum, and this effect is
described by the corresponding component $D_{ij}$ of the
deformation potential. In the above analysis of the simplest model
it has been implied that $u_{ij}$ can be represented in the form $u_{ij}=uu_{ij}^0$ where $u_{ij}^0$ are some constants, and $u$ describes the magnitude of the deformation. Then, we  arrive at the problem considered above with $D_{s,l}= \sum_{i,j} D_{ij}^{s,l}u_{ij}^0$ and $C=\sum_{i,j,l,m} c_{ijlm}u_{ij}^0u_{lm}^0$ where $c_{ijlm}$ are the elastic moduli of the crystal. However, the constants $u_{ij}^0$, in general, can change at the transition. In
particular, if a metal contains several equivalent groups of
charge carriers, in minimizing the appropriate energy it may be favorable to break a symmetry of these groups. In other words, one should minimize the energy over all $u_{ij}$ independently, and this minimization can lead to the crystal symmetry breaking. To illustrate this idea, consider first a model spectrum imitating the band structure of bismuth, and then we
shall generalize the obtained results.

\subsection{Model imitating bismuth}\label{IVa}

\begin{figure}[t] 
\includegraphics{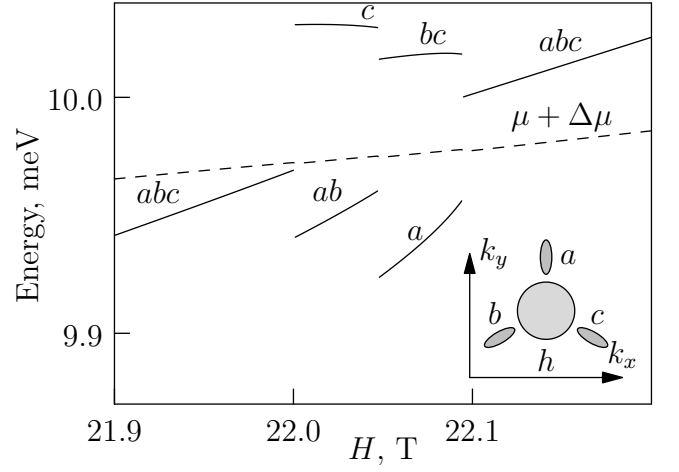}
\caption{\label{fig2} The Landau-level edge $\epsilon_e^1(0)+ \Delta\varepsilon_e^i$ for
the electron ellipsoids ``a'', ``b'', and ``c'' versus $H$ (solid
lines). The dashed line is the $H$-dependence of $\mu+\Delta \mu$.
All the energies are measured from $\varepsilon_e$, the bottom of
the undeformed electron ellipsoids. Here $A=3\cdot10^{17}$
cm$^{-3}$meV$^{-1}$, $B/A=0.5$, $A_h=-0.4$, $e\sqrt{2m_s}
/(3\pi^2\hbar^2cA)=0.04$ meV$^{1/2}$/T, $\mu=10$ meV, $m_\perp
=0.2m$, $g_e(m_\perp/4m)=0.4$. These parameters gives $H_1\approx
20.2$ T. The inset schematically shows the 3 electron ellipsoids
centered at the points L and the hole ellipsoid (h) at the point T
of the Brillouin zone.
 } \end{figure}

Let the Fermi surface of a metal with the symmetry of bismuth
\cite{Ed}  consist of three equivalent  electron ellipsoids
``a'', ``b'', ``c'' centered at the points L of the Brillouin zone
and of a large ellipsoid (similar to the large sphere in the
simplest model) located at the point T, Fig~\ref{fig2}. The large
ellipsoid can be of the hole or electron type, this has no effect
on the subsequent results. The axes 1 and 3 coincide with the
binary and the trigonal axes, respectively, while the axis 2 is
along the bisectrix direction. The spectra of the electrons,
$\epsilon_e({\bf k})$, and of the charge carriers in the large
group, which will be arbitrarily called the ``holes'',
$\epsilon_h({\bf k})$, are assumed to be quadratic functions of
${\bf k}$. In particular, we use the following dispersion relation
for the electrons:
 \begin{eqnarray}\label{15}
 \epsilon_e({\bf k})=\varepsilon_e+\frac{k_1^2+k_2^2}{2m_{\perp}}+
 \frac{k_3^2}{2m_s},
 \end{eqnarray}
where $\varepsilon_e$ is the bottom of the electron bands, and we
admit a difference between the effective masses $m_s$ and
$m_{\perp}$.

The elastic energy $E_{el}$ for such a crystal is the quadratic
form in $u_{ij}$: \cite{Mi}
 \begin{eqnarray}\label{16a}
&E_{el}&\!\!\!\!=\frac{c_{11}+c_{12}}{4}(u_{11}+u_{22})^2 +
\frac{c_{33}}{2}u_{33}^2  \\
&+&\!\!\!\!\frac{c_{11}-c_{12}}{4}[(u_{11}-u_{22})^2 +4u_{12}^2]
+c_{13}(u_{11}+u_{22})u_{33}\nonumber
\\ &+&\!\!\!\!2c_{44}(u_{13}^2+u_{23}^2)\!+\!
2c_{14}[(u_{11}-u_{22})u_{23}\!+\!2u_{12}u_{13}], \nonumber
 \end{eqnarray}
where $c_{11}$, $c_{12}$, $c_{33}$, $c_{13}$, $c_{14}$, and
$c_{44}$ are the elastic moduli of the crystal in the Voigt
notation. The deformations $u_{ij}$ shift the energy extremum for
the holes, $\varepsilon_h$, and the bottom of the electron band,
$\varepsilon_e$, in the ellipsoids ``a'', ``b'', ``c'' as follows:
\cite{Mi}
 \begin{eqnarray}\label{17a}
\Delta \varepsilon_h\!\!\!&=&\!\!D_{11}^h(u_{11}+u_{22}) +
D_{33}^hu_{33},\\
\Delta\varepsilon_e^a\!\!\!&=&\!\!D_{11}^eu_{11}+D_{22}^eu_{22}+
D_{33}^eu_{33}+ 2D_{23}^eu_{23},\label{18a}
\\ \Delta \varepsilon_e^{b,c}\!\!\!&=&\!\!\frac{1}{4}
(D_{11}^e\!+3D_{22}^e)u_{11}\!+\frac{1}{4}(3D_{11}^e\!+D_{22}^e)
u_{22}\!+ D_{33}^eu_{33}\nonumber
\\ \!\!\!&\pm& \!\!\frac{\sqrt{3}}{2}(D_{11}^e\!-D_{22}^e)u_{12}\!
\pm \sqrt{3}D_{23}^eu_{13}\!-D_{23}^eu_{23},\label{19a}
 \end{eqnarray}
where $D_{ij}^e$, $D_{ij}^h$ are the components of the deformation
potential for the electrons and the holes, respectively.

Let the magnetic field $H$ be along the trigonal axis of the
crystal. In this case the electron Landau levels at $k_3=0$ has
the form:
 \begin{eqnarray}\label{2}
\epsilon_e(n,\pm,0)=\varepsilon_e+\frac{\hbar eH}{m_\perp c}\left
(n+\frac{1}{2} \pm g_e\frac{m_\perp}{4m}\right ),
 \end{eqnarray}
where $g_e$ is the electron $g$ factor, and $m$ is the electron
mass. We assume here that the lowest electron Landau level $0_e^-$
is filled, while the next Landau level of the electrons, $0_e^+$
is close to the Fermi energy ($0$ means $n=0$, and the minus and
plus indicate the projection of the electron spin on the direction of $H$). In strong magnetic fields ($H>10$ T) this situation really occurs in bismuth in a wide interval of the magnetic fields. \cite{SM1} The magnetostriction of the metal is still found from the minimization of $E(u_{ij},H)$, Eq.~(\ref{1a}), with respect to all $u_{ij}$. But now the term $Cu^2/2$ is replaced by $E_{el}(u_{ij})$,
Eq.~(\ref{16a}), whereas $E_e$ is the sum of the energies of the
holes and of the electrons in the ellipsoids ``a'', ``b'', and
``c'', $E_e=E_h+E_a+E_b+E_c$. The changes of these energies are
described by the formulas that are similar to Eqs.~(\ref{2a}) and
(\ref{5a}).

Using Eqs.~(\ref{17a})--(\ref{19a}), it is convenient to express
four components of the tensor $u_{ij}$ in terms of $\Delta
\varepsilon_h$, $\Delta \varepsilon_e^a$, $\Delta
\varepsilon_e^b$, $\Delta \varepsilon_e^c$ and to insert these
expressions into $E(u_{ij},H)$. The energy $E$ thus obtained is a
quadratic form in the remaining two components of $u_{ij}$ and in
$\Delta \varepsilon_h$, and we minimize this form with respect to
these three variables. Eventually, we arrive at:
 \begin{eqnarray}\label{20a}
&E&\!\!\!(\Delta \varepsilon_e^a,\Delta \varepsilon_e^b,\Delta
\varepsilon_e^c,H)= A[(\delta \varepsilon_e^a)^2 +
(\delta\varepsilon_e^b)^2 + (\delta \varepsilon_e^c)^2]\nonumber
\\ &+&2B[\delta \varepsilon_e^a\,\delta \varepsilon_e^b+ \delta
\varepsilon_e^b\,\delta\varepsilon_e^c + \delta
\varepsilon_e^c\,\delta \varepsilon_e^a]+
(\Omega_a^1+\Omega_b^1+\Omega_c^1) \nonumber \\
&+&[N_e^0(H)-N_e(0)](\delta \varepsilon_e^a+\delta
\varepsilon_e^b+\delta \varepsilon_e^c),
\end{eqnarray}
where $\delta \varepsilon_e^i\equiv \Delta
\varepsilon_e^i-\Delta\varepsilon_h$ ($i=a,b,c$); $\Delta
\varepsilon_h=A_h (\Delta \varepsilon_e^a+\Delta
\varepsilon_e^b+\Delta \varepsilon_e^c)$; the coefficients $A$,
$B$, and $A_h$ are combinations of the elastic moduli and the
components of the deformation potential, and the explicit
expressions for them are given in the Appendix; $N_e(0)$ is the
number of the electrons in one of the {\it undeformed} ellipsoids
at $H=0$, and $N_e^0(H)$ denotes the number of the electrons in
the lowest Landau level $0_e^-$ of this ellipsoid; $\Omega_i^1$
are the $\Omega$ potentials for the electrons occupying the next
Landau level $0_e^+$ in the {\it deformed} ellipsoids $i=$ a, b,
c. These potentials $\Omega_i^1$ are determined  by
Eq.~(\ref{6a}):
 \begin{equation}\label{21}
\Omega_i^1\!=\!\frac{1}{2}\Omega_s^1(\mu\!+\Delta
\mu\!-\!\epsilon_e^1(0)-\!\Delta \varepsilon_e^i)\!=\!\frac{1}{2}
\Omega_s^1(\mu\!-\!\epsilon_e^1(0)\!-\! \delta \varepsilon_e^i),
 \end{equation}
where $\epsilon_e^1(0)$ is the edge of the Landau level $0_e^+$,
\[
 \epsilon_e^1(0)\equiv \epsilon_e(0,+,0)=\varepsilon_e+
 \frac{\hbar eH}{m_\perp c}\left[\frac{1}{2}+
 g_e\frac{m_\perp}{4m}\right].
\]
As in the simplest model, the Fermi energy $\mu(H)$ and its shift
$\Delta \mu$ under the deformations are found from the relations:
$\mu(H)\approx \mu(0)\equiv \mu$ and $\Delta \mu \approx
\Delta\varepsilon_{h}$, and these relations have been used in the
derivation of Eq.~(\ref{20a}).

Let us introduce the characteristic magnetic field $H_1$ at which
$\epsilon_e^1(0)$ crosses the Fermi energy $\mu$. When $H$ is
far from $H_1$, i.e., when the edge of the Landau level $0_e^+$ is
not close to the Fermi energy, the terms $\Omega_i^1$ is practically  linear in $\delta\varepsilon_e^i$:
$\Omega_i^1\approx \delta\varepsilon_e^iN_e^1(H)$ where $N_e^1(H)$ is numbers of the electrons in the Landau level $0_e^+$ of one of the
undeformed ellipsoids. Then, expression (\ref{20a}) reduces to
the quadratic form in $\delta \varepsilon_e^i$. The minimization of this expression with respect to $\delta \varepsilon_e^i$ always leads to $\delta \varepsilon_e^a=\delta \varepsilon_e^b =\delta \varepsilon_e^c$, i.e., the magnetostriction does not change the crystal symmetry of the metal at such magnetic fields, and these $\delta \varepsilon_e^i$ are smooth functions of $H$. In
the case when the Landau-level edge $\epsilon_e^1(0)$ is close to
the Fermi energy, $\Omega_i^1$ become nonlinear and
singular functions of $\delta\varepsilon_e^i$, and $\Delta
\varepsilon_e^i$ exhibit a jump. Moreover, at the additional condition $B>0$ the symmetry breaking appears. Figure \ref{fig2} shows the results of the minimization of Eq.~(\ref{20a}) in this case for the parameters comparable with the parameters of bismuth (see Appendix). It is seen that in some interval of $H$, the
shifts $\Delta\varepsilon_e^i$ become different, and the Landau
level $0_e^+$ is not the same for the three ellipsoids. In other
words, the trigonal symmetry  of the ellipsoids and of the
magnetostriction breaks in this interval of $H$, and
the successive three phase transitions are visible in Fig.~\ref{fig2}. We emphasize that this symmetry breaking is due to the singular terms $\Omega_i^1$ producing minima of $E(\Delta \varepsilon_e^a,\Delta
\varepsilon_e^b,\Delta \varepsilon_e^c,H)$ at $\Delta
\varepsilon_e^i$ which do not coincide with each other.

We now briefly discuss the experimental results of
Refs.~\onlinecite{B2,KBstrangePeaks,behnia,behnia11}. In contrast
with the model used here, realistic models for the spectrum of
bismuth reveal proximity of $0_e^+$ and $\mu$ in a wide interval
of the magnetic fields. \cite{SM1,Zhu} Then, the interval between
the fields of the transitions can increase essentially, and it is
not improbable that the unusual peaks observed in the Nernst
coefficient of bismuth \cite{B2,KBstrangePeaks}  correspond to
these transitions. As to the asymmetry of the angular dependence
of the magnetostriction,\cite{behnia} its existence, at least for
not-too-large tilt angles $\theta$, can be explained by the
obtained spontaneous symmetry breaking of the electron ellipsoids.
In experiments of Ref.~\onlinecite{behnia11} the
magnetoresistivity was measured for magnetic fields lying in the
basal plane that is  perpendicular to the trigonal axis. For such
magnetic fields the spontaneous symmetry breaking of two electron
ellipsoids can occur when the magnetic field is along the
bisectrix direction. In principle, this breaking can lead to an asymmetric dependence of magnetoresistivity on the angle $\phi$ defining the direction of the magnetic field in the basal plane.\cite{behnia11} Of course, these qualitative considerations require additional theoretical and experimental investigations.

\subsection{Some generalizations}\label{IVb}

The approach used in Sec.~\ref{IVa} can be easily extended to the
case of multivalley metals with a large electron group and with
$n$ equivalent electron pockets. Existence of the large electron group is exclusively assumed to simplify the equation determining the Fermi level. Due to the time reversal symmetry connecting the points ${\bf k}$ and ${-\bf k}$ in the Brillouin zone, the relations ($i=1,\dots,n$)
\begin{equation}\label{41}
 \Delta \varepsilon_i=\sum_{jj'} D_{jj'}^i u_{jj'}
 \end{equation}
between the energy-extrema shifts $\Delta \varepsilon_i$ of
the equivalent electron pockets and the deformation tensor
$u_{jj'}$ are the same for the pockets at ${\bf k}$ and ${-\bf
k}$. (Here $D_{jj'}^i$ is the deformation potential.) Thus, the
number $\tilde n$ of the independent relations (\ref{41})
for the equivalent pockets is $n/2$ if the pockets are inside the
Brillouin zone and is equal to $n$ if they are on its faces.
Possible values of $\tilde n$ ($\tilde n=1-4, 6$) do not exceed
$6$, the number of the component $u_{jj'}$ of the deformation
tensor. Using these independent relations and a similar equality
for the large electron group, one can express a part of $u_{jj'}$
(or even all $u_{jj'}$ if $\tilde n=6$) in terms of $\Delta
\varepsilon_i$ where $i=1,\dots,\tilde n+1$, and the index $\tilde n+1$ marks the large electron group. Expressing the energy $E(u_{jj'},H)$ via these $\Delta \varepsilon_i$, and minimizing the energy $E(\Delta\varepsilon_{i},H)$ thus obtained with respect to the other $u_{jj'}$ and to $\Delta \varepsilon_{\tilde n+1}$ (if $\tilde n <6$), we arrive at the formula similar to Eq.~(\ref{20a}):
 \begin{eqnarray}\label{42}
\!\!\!&E&\!\!\!(\delta \varepsilon_i,H)=A'\sum_{i=1}^{\tilde
n}(\delta \varepsilon_i)^2+2B'\!\!\!\!\!\sum_{i,j=1,i<j}^{\tilde
n}\delta \varepsilon_i\delta \varepsilon_j\nonumber \\
\!\!\!&+&\!\!\frac{n}{\tilde n}\sum_{i=1}^{\tilde
n}\Omega_i^1(\mu-\!\epsilon^1\!\!-\delta\varepsilon_i)\!
+\![N^0(H)\!-\!N(0)]\frac{n}{\tilde n}\sum_{i=1}^{\tilde n}\delta
\varepsilon_i,\ \ \ \ \
\end{eqnarray}
where $\delta\varepsilon_i=\Delta\varepsilon_i -\Delta\varepsilon_{\tilde n+1}$, $\Delta \varepsilon_{\tilde n+1} =A_{\tilde n+1} \sum_{i=1}^{\tilde n}\Delta
\varepsilon_i$; the coefficients $A'$, $B'$, and $A_{\tilde n+1}$ are combinations of the elastic moduli of the metal and of its
components of the deformation potential, $N(0)$ is the numbers of
the electrons in one of the {\it undeformed} pockets at $H=0$, and
$N^0(H)$ denotes the number of the electrons in all the Landau levels of this pocket except the Landau subband nearest to $\mu$;
$\epsilon^1$ is the edge of this subband; $\Omega_i^1$ are the
$\Omega$ potentials for the electrons occupying the Landau level
nearest to $\mu$ in the {\it deformed} pockets $i=1, \dots, \tilde n$. Since $\mu-\epsilon^1$ is usually assumed to be small, in the Landau subband nearest to $\mu$ the dependence of the electron energy on the wave vector $k_3$ parallel to the magnetic field can be
approximated by $\hbar^2k_3^2/2m_s$ with some mass $m_s$. Then, one
can use formula (\ref{6a}) in the calculation of $\Omega_i^1$. In
the derivation of Eq.~(\ref{42}) we have used that the Fermi
energy $\mu(H)$ and its shift $\Delta \mu$ under the deformations
are found from the relations: $\mu(H)\approx \mu(0)\equiv \mu$
and $\Delta \mu \approx \Delta \varepsilon_{\tilde n+1}$. As a result of these relations, the energy $E$ depends on the
differences $\delta\varepsilon_i$ only. In the case $\tilde n=6$,
we do not minimize $E(\Delta\varepsilon_{i},H)$ with respect to
$\Delta \varepsilon_{\tilde n+1}$. Now $\Delta \varepsilon_{\tilde n+1}$ is directly expressed via $\Delta \varepsilon_{1}\dots \Delta \varepsilon_{\tilde n}$ with formula (\ref{41}) for $i= n+1$. However, eventually we still arrive at the same formula (\ref{42}) but with other expressions for $A_{\tilde n+1}$, $A'$, and $B'$.

In Secs.~\ref{II} and \ref{IVa}, in fact, we have considered the
cases $\tilde n=1$ and $3$. It is also instructive to consider the
case $\tilde n=2$ since this case shed light on existence of the
symmetry breaking transitions for all even values of $\tilde n$, $\tilde n=2, 4, 6$. Indeed, for even $\tilde n$ the set of the electron pockets always can be divided into two equal parts, and we may put $\Delta \varepsilon_i=\Delta \varepsilon_1$ for the first part and $\Delta \varepsilon_i=\Delta \varepsilon_2$ for the second part. If the transition occurs under these constraints, it also occurs in situation when all $\Delta \varepsilon_i$ can be different. In other words, the case $\tilde n=2$ gives sufficient conditions for the symmetry breaking at even $\tilde n$.

 \begin{figure}[t] 
\includegraphics{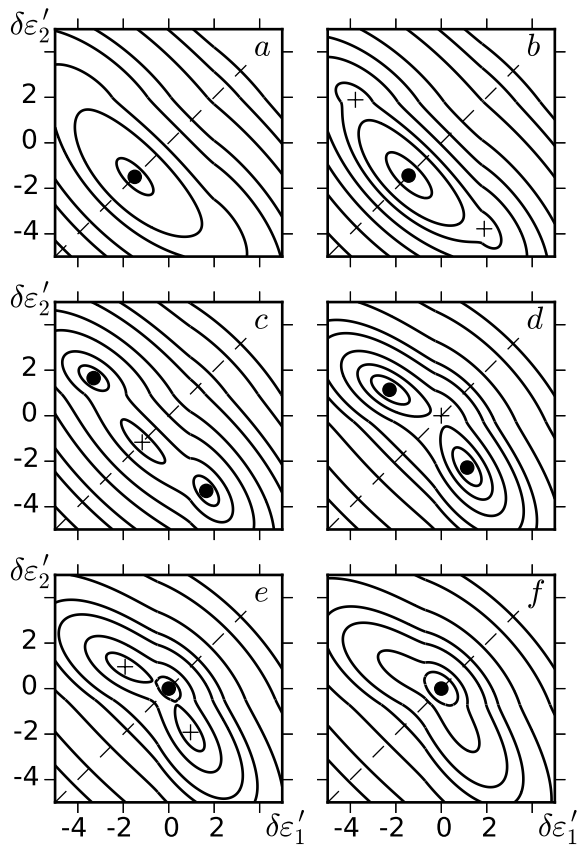}
\caption{\label{fig3} The lines of constant $E(\delta
\varepsilon_1,\delta \varepsilon_2,H)$ in the plane
$\delta\varepsilon_1' $--$\delta\varepsilon_2'$  under the
condition that the Landau-level edge $\epsilon^1$ of the two
equivalent pockets is close to the Fermi energy $\mu$. The black solid dots mark the absolute minima of the energy $E$, whereas the crosses mark its local minima. Here $\delta\varepsilon_i'\equiv \delta\varepsilon_i-\delta\varepsilon_{\rm sym}$ are measured in meV, and $E$ is defined by Eq.~(\ref{42}) with $\tilde n=2$, $B'/A'=0.5$,
$e\sqrt{2m_s}H_1 /(3\pi^2\hbar^2cA)=0.8$ meV$^{1/2}$. The magnetic
field increases from ``a'' to ``f'' and $\mu-\epsilon^1=0.5$ meV
(a), $0.25$ meV (b), $-0.15$ meV (c), $-0.25$ meV (d), $-0.3$ meV
(e), $-0.6$ meV (f).
 } \end{figure}   

For $\tilde n=2$ the first, second, and forth terms in
Eq.~(\ref{42}) produce a minimum of $E$ in the plane
$\delta\varepsilon_1$--$\delta\varepsilon_2$. This minimum lies in
the straight line $\delta\varepsilon_1=\delta\varepsilon_2$ and
occurs at the point
\[
\delta\varepsilon_1=\delta\varepsilon_2=
-\frac{[N^0(H)\!-\!N(0)]}{2(A'+B')}\frac{n}{\tilde n}\equiv
\delta\varepsilon_{\rm sym}.
\]
The singular terms associated with $\Omega_i^1(\mu-\epsilon^1
-\delta\varepsilon_i)$ not only can shift this minimum but also
can cause additional minima that do not necessarily lie in the
line $\delta\varepsilon_1=\delta\varepsilon_2$. In other words,
they can lead to the spontaneous symmetry breaking of the electron
pockets, Fig.~\ref{fig3}. In Fig.~\ref{fig3} we show the lines of
constant $E$ in the plane
$\delta\varepsilon_1'$--$\delta\varepsilon_2'$ where
$\delta\varepsilon_1'\equiv
\delta\varepsilon_1-\delta\varepsilon_{\rm sym}$ and
$\delta\varepsilon_2'\equiv
\delta\varepsilon_2-\delta\varepsilon_{\rm sym}$. According to
Fig.~\ref{fig3}, the spontaneous symmetry breaking occurs at a
magnetic field $H_1^-$ that corresponds to the panel ``c'',
whereas the electron pockets return to the symmetric state at a
field $H_1^+$ that corresponds to the panel ``e''. The fields
$H_1^-$ and $H_1^+$ are close to the field $H_1$ at which
$\epsilon^1$ crosses the Fermi energy, $\epsilon^1(H_1)=\mu$,
while their difference can be estimated as follows:
 \begin{equation}\label{43}
(H_1^+\!-\!H_1^-)\frac{e\hbar}{m_{\perp}c}=\frac{27\alpha^2B'}{64
(A'-B')^2(A'+B')},
 \end{equation}
where $m_{\perp}$ is the cyclotron mass of the pockets at $k_3=0$,
and $\alpha=-(2eH_1\sqrt{2m_s})/(3\pi^2\hbar^2c)$ has the same
form as in Eq.~(\ref{10}). The symmetry breaking occurs only at
$B'>0$ (the conditions $A'>0$ and $|B'|/A'<1$ follow from
positivity of the elastic energy). At $B'<0$ the phase transition
occurs without the symmetry breaking, i.e., this transition is
analogous to that considered in Sec.~\ref{II}. Since the constants
$A'$ and $B'$ are similar to $A$ and $B$ given in the Appendix,
they can be estimated as $A'\sim B'\sim C/D^2$ where $C$ and $D$
are characteristic values of the elastic moduli and of the
deformation potential in the metal. If we put
$H_1(e\hbar/m_{\perp}c) \sim \mu-\varepsilon_1$ where
$\varepsilon_1$ is the energy-band edge for one of the equivalent
pockets, we obtain the estimate for $(H_1^+\!-\!H_1^-)/H_1$,
 \[
\frac{H_1^+\!-\!H_1^-}{H_1} \sim \frac{\alpha^2 D^4}
{C^2(\mu-\varepsilon_1)},
 \]
which is of the order of the value given by Eq.~(\ref{13}).

Compare now the above symmetry breaking that is characteristic of metals with the Jahn-Teller effect \cite{JT} which occurs in molecules and solid insulators. When none of the Landau-subbands edges is close to the  Fermi energy, the terms $\Omega_i^1$ in formula (\ref{42}) vanish or become linear in $\delta\varepsilon_i$. Then, expression (\ref{42}) reduces to the positively definite quadratic form in $\delta\varepsilon_i'$:
 \begin{equation}\label{44}
 E(\delta \varepsilon_i',H)=A'[(\delta \varepsilon_1')^2 + (\delta \varepsilon_2')^2]+ 2B'\delta \varepsilon_1'\delta \varepsilon_2'.
 \end{equation}
The minimization of this form yields $\delta \varepsilon_1'=\delta \varepsilon_1'=0$, which means the absence of asymmetric deformations in this case. In the theory of the Jahn-Teller effect \cite{JT} a   quadratic form for the appropriate energy contains linear terms, and these terms shift the energy minimum from the origin of the coordinate, i.e., lead to the spontaneous symmetry breaking.
In the metals, when a Landau-subband edge is close to the Fermi energy, the {\it singular} terms $\Omega_i^1$ are added to the right hand side of Eq.~(\ref{44}), and the symmetry breaking is due to these terms which have no counterparts in the theory of the Jahn-Teller effect.

At the end of this section we briefly discuss the situation when
the large electron group is absent, and a metal contains only two
equivalent electron pockets ``a'' and ``b''. In this case the
Fermi level calculated from the conservation of the electrons:
$N_a+N_b=$const., has a complicated dependence on the magnetic
field and on the deformations. Nevertheless, our numerical calculations show that the appropriate results do not differ qualitatively from those presented above. In particular, the difference $(H_1^+\!-\!H_1^-)$ at the same values of the parameters is comparable with that given by Eq.~(\ref{43}). For this reason, in
this paper we have assumed existence of the large electron group,
which simplifies our analysis.

\section{Conclusions}\label{V}

We  have shown that in strong magnetic fields the first-order phase transitions can take place in metals with a multivalley band structure. These transitions are caused by the electron-phonon interaction, and they occur when an electron Landau level approaches   the Fermi energy of the metals. At the transitions the magnetostriction and the magnetization experience jumps.

In metals with several equivalent groups of charge carriers, in certain intervals of magnetic fields a spontaneous symmetry breaking of the magnetostriction can occur that changes a crystal symmetry of the metals. These intervals are in the vicinities of the magnetic fields at which Landau levels of the equivalent groups are close to the Fermi energy. This result reveals possibility of governing the crystal symmetry of the metals by a magnetic field.

\appendix
\section{Parameters of the model imitating the band
structure of bismuth}\label{app}

Direct calculations described in Sec.~\ref{IVa} give the following
expressions for the coefficients $A$, $B$, and $A_h$:
\begin{eqnarray}
A&=&\frac{\tilde{A}}{D_A}+2\frac{\tilde{B}}{D_B}, \\
B&=&\frac{\tilde{A}}{D_A}-\frac{\tilde{B}}{D_B}, \label{A2}
\\ A_h&=&-\frac{F_1}{3F_2},
\end{eqnarray}
where
\begin{equation}
\tilde{A}=\frac{c_{33}(c_{11}+c_{12})-2c_{13}^2}{36},
\end{equation}
\begin{eqnarray}
D_A&=&c_{33} [D_{11}^h-0.5(D_{11}^e+D_{22}^e)]^2\nonumber \\
&-&2c_{13}[D_{11}^h-0.5(D_{11}^e+
D_{22}^e)][D_{33}^h-D_{33}^e]\nonumber \\ &+&
0.5(c_{11}+c_{12})[D_{33}^h- D_{33}^e]^2,
\end{eqnarray}
\begin{equation}
\tilde{B}= \frac{2(c_{11}-c_{12})c_{44}-4c_{14}^{2}}{9},
\end{equation}
\begin{eqnarray}
D_B&=& c_{44}(D_{11}^{e}-
D_{22}^{e})^{2}+2(c_{11}-c_{12})(D_{23}^{e})^{2}\nonumber \\&-&
4c_{14}D_{23}^{e}(D_{11}^{e}-D_{22}^{e}),
\end{eqnarray}
\begin{eqnarray}
F_1&=&c_{33}D_{11}^h\!
\left(D_{11}^h-\frac{D_{11}^e+D_{22}^e}{2}\right)\! \nonumber \\
&-&\!c_{13}\!\left[D_{11}^h(D_{33}^h-D_{33}^e)+D_{33}^h
\left(D_{11}^h- \frac{D_{11}^e+D_{22}^e}{2}\right)\right]\!
\nonumber \\&+& \!\frac{c_{11}+c_{12}}{2}D_{33}^h(D_{33}^h-
D_{33}^e),
\end{eqnarray}
\begin{eqnarray}
&F_2&\!\!\!=-c_{33}\frac{D_{11}^e+D_{22}^e}{2}\!
\left(D_{11}^h-\frac{D_{11}^e+D_{22}^e}{2}\right)\nonumber
\\&+&\!\!\!c_{13}\! \Big[\frac{D_{11}^e\!+\!D_{22}^e}{2}
(D_{33}^h\!-\!D_{33}^e)\!+\!D_{33}^e\!\left(\!D_{11}^h\!-
\!\frac{D_{11}^e\!+\!D_{22}^e}{2}\!\right)\!\Big ]\nonumber \\
&-&\frac{c_{11}+c_{12}}{2}D_{33}^e(D_{33}^h-D_{33}^e).
\end{eqnarray}

According to Hansen et al., \cite{Han} one has 1 eV $\lesssim
|D_{ij}^e|, |D_{ij}^h| \lesssim $ 8 eV for bismuth. On the other
hand, $c_{ij}\sim (7-64)\cdot 10^{10}$ erg/cm$^3$ in this
material. \cite{Bret} With these values of $D_{ij}$ and $c_{ij}$,
we take the following values of the parameters for our
calculations of the Landau levels presented in Fig.~2: $A=3\cdot
10^{17}$ cm$^{-3}$meV$^{-1}$, $B/A=0.5$, $A_h=-0.4$.


\begin{thebibliography}{}

\bibitem{B1} K. Behnia, M.A. Measson, and Y. Kopelevich, \prl{\bf
98}, 166602 (2007).

\bibitem{SM1} Yu. V. Sharlai and G. P. Mikitik, \prb{\bf 79},
081102(R) (2009).

\bibitem{AB} J. Alicea and L. Balents, \prb{\bf79}, 241101 (2009).

\bibitem{SM2} Yu. V. Sharlai and G. P. Mikitik, \prb{\bf 83},
085103 (2011).

\bibitem{B2} K. Behnia, L. Balicas, and Y. Kopelevich,  Science
{\bf 317}, 1729 (2007).

\bibitem{KBstrangePeaks} H. Yang, B. Fauque, L. Malone,
A.B. Antunes, Z. Zhu, C. Uher, and K. Behnia, Nature Comm. {\bf
1}, 47 (2010).

\bibitem{Ong} Lu Li, J.G. Checkelsky, Y.S. Hor, C. Uher,
A.F. Hebard, R.J. Cava, N.P. Ong, Science {\bf 321}, 547 (2008).

\bibitem{Ser} B. Seradjeh, J. Wu, P. Phillips, \prl{\bf
103}, 136803 (2009).

\bibitem{B3} K. Behnia, \prl{\bf 104}, 059705 (2010).

\bibitem{Ser1} B. Seradjeh, J. Wu, P. Phillips, \prl{\bf
104}, 059706 (2010).

\bibitem{Zhu} Z. Zhu, B. Fauque, L. Malone, A.B. Antunes,
Y. Fuseya, and K. Behnia, Proc.\ Natl.\ Acad.\ Sci.\ U.S.A.\ {\bf 109}, 14813 (2012).

\bibitem{ZhuQuestions} The volume of the domain did not
correspond to magnitude of the experimental signal. Besides, in
Ref.~\onlinecite{Zhu} the chemical potentials in the different
domains were assumed to be different.

\bibitem{Sh} D. Shoenberg, {\it Magnetic Oscillations in Metals}
(Cambridge University Press, Cambridge, England, 1984).

\bibitem{behnia11} Z. Zhu, A. Collaudin, B. Fauque, W. Kang and
K. Behnia, Nature Phys. {\bf 8}, 89 (2012).

\bibitem{behnia} R. Kuchler, L. Steinke, R. Daou, M. Brando,
K. Behnia, and F. Steglich, Nature Materials {\bf 13}, 461 (2014).

\bibitem{abanin} D. A. Abanin, S. A. Parameswaran, S. A. Kivelson
and S. L. Sondhi, \prb{\bf 82}, 035428 (2010).

\bibitem{JT} R.S. Knox, A. Gold, {\it Symmetry in the Solid State}
(W.A. Benjamin, Inc., New York, Amsterdam, 1964).

\bibitem{Kap} P. Kapitza, Proc. R. Soc. London, Ser. A {\bf 135},
568 (1932).

\bibitem{Han} O.P. Hansen, I.F.I. Mikhail, M. Yu. Lavrenyuk, N.
Ya. Minina, J. Low Temp. Phys. {\bf 95}, 481 (1994).

\bibitem{Mi} J.-P. Michenaud, J. Heremans, M. Shayegan,
C. Haumont, \prb{\bf 26}, 2552 (1982).

\bibitem{Littlewood} P.B. Littlewood, B. Mihaila, R.C. Albers,
\prb{\bf 81}, 144421  (2010).

\bibitem{com} The difference in the signs is due to incompleteness
of formula (4) used in Ref.~\onlinecite{Littlewood} for the energy
of electrons in a magnetic field. In this formula the contribution
$(eHV\sqrt{2m}/\pi^2c\hbar^2) \sum_r(r+0.5)\beta
H[\zeta-(r+0.5)\beta H]^{1/2}$ (in the notations of
Ref.~\onlinecite{Sh}) was lost; cf. this formula with Eq.~(14) in
the second Appendix of Ref.~\onlinecite{Sh}.

\bibitem{Ed} V.S. Edel'man, Usp. Fiz. Nauk {\bf 123}, 257 (1977)
[Sov. Phys. Usp. {\bf 20}, 819 (1977)].

\bibitem{Bret} A. de Bretteville, Jr., E.R. Cohen, A.D. Balatto,
I.N.  Greenberg, Phys.\ Rev.\ {\bf 148}, 575 (1966).


\end{thebibliography}
\end{document}